\input{aipcheck}

\documentclass[
    ,final            % use final for the camera ready runs
  ]
  {aipproc}

\layoutstyle{6x9}

\newcommand{\nlsim}{\mathrel{\rlap{\lower4pt\hbox{\hskip0pt$\sim$}} 
 \raise1pt\hbox{$<$}}}           %less than or approx. symbol 
\newcommand{\ngsim}{\mathrel{\rlap{\lower4pt\hbox{\hskip0pt$\sim$}} 
 \raise1pt\hbox{$>$}}}           %greater than or approx. symbol 

\def\gtorder{\mathrel{\raise.3ex\hbox{$>$}\mkern-14mu
 \lower0.6ex\hbox{$\sim$}}}
\def\ltorder{\mathrel{\raise.3ex\hbox{$<$}\mkern-14mu
 \lower0.6ex\hbox{$\sim$}}}

\def\gegm{G_E / G_M}
\def\ge{G_E}
\def\gm{G_M}

\begin{document}

\title{Two-photon exchange measurements with positrons and electrons}


%13.40.Gp Electromagnetic form factors 
%13.40.Ks Electromagnetic corrections to strong- and weak-interaction processes 
%25.30.Bf Elastic electron scattering 
%25.30.Hm Positron-induced reactions 

%13.60.Fz Elastic and Compton scattering 
%13.60.-r Photon and charged-lepton interactions with hadrons (for neutrino interactions, see 13.15.+g)
%13.88.+e Polarization in interactions and scattering 


\author{John Arrington}{
  address={Physics Division, Argonne National Laboratory, Argonne, IL 60439}
}

\begin{abstract}

Two-photon exchange contributions have potentially broad ranging impact on
several charged lepton scattering measurements.  Previously believed to
be extremely small, based in part on comparisons of positron scattering
and electron scattering in the 1950s and 1960s, recent data suggest that
the corrections may be larger than expected, in particular in kinematic
regions that were inaccessible in these early positron scattering measurements.
Additional measurements using positron beams at Jefferson Lab would allow for
a detailed investigation of these contributions in a range of reactions
and observables.

\end{abstract}

\maketitle

%%%%%%%%%%%%%%%%%%%%%%%%%%%%%%%%%%%%%%%%%%%%
%% MAINMATTER
%%%%%%%%%%%%%%%%%%%%%%%%%%%%%%%%%%%%%%%%%%%%

\section{Introduction}

The nucleon electromagnetic form factors are fundamental quantities that relate
to the charge and magnetization distributions in the
nucleon~\cite{arrington07a, perdrisat07, miller03, kelly02}.  Thus, they are
important quantities when examining the spatial distribution and dynamics of
quarks in the nucleon ~\cite{miller07, miller08b, miller09a}.  Unpolarized
elastic scattering has been used since the 1950s to obtain the proton electric
and magnetic form factors, $\ge$ and $\gm$, using the Rosenbluth separation
technique~\cite{rosenbluth50}.  In certain kinematic regions, it is difficult
to separate the electric and magnetic contributions to the cross section, and
in particular it is difficult to extract $\ge$ from the cross sections at high
$Q^2$ values~\cite{andivahis94,walker94,christy04}.
Polarization measurements have an important role in overcoming
this limitation, as they are sensitive to the ratio $\gegm$~\cite{akhiezer68,
dombey69}.  The $Q^2$ measurements at Jefferson Lab~\cite{jones00, gayou02}
showed a striking disagreement with previous measurements~\cite{arrington03a},
as well as a new, high precision extraction using a modified Rosenbluth
separation technique~\cite{qattan05}.

Early speculation was that the discrepancy could be the result of two-photon
exchange (TPE) contributions, which are neglected (except for IR-divergent
contributions) in standard radiative correction procedures based on the
formalism of Mo and Tsai~\cite{mo69}.  Estimates of contributions beyond
the IR divergent terms suggested that any additional effects were
small~\cite{greenhut69}, and this was supported by comparisons of positron and
electron scattering~\cite{mar68} where the TPE contributions change sign.
More recently, the TPE contributions to the unpolarized cross section were
reexamined~\cite{maximon00}, and it was also shown that these contributions
could potentially have a large impact on the Rosenbluth extractions while
having little impact on the polarization observables~\cite{guichon03}.

Since then, several approaches have been used to calculate the TPE
contributions~\cite{blunden05a, afanasev05a, borisyuk06b, borisyuk08, jain08,
kuhn08, arrington04c}, for both cross section and polarization observables, as
well as examination of other reactions or observables~\cite{dong06,
kondratyuk06, afanasev05b, rekalo04, arrington07b, tjon09}. The
cross section calculations have significant model dependence, but generally
agree on the qualitative features~\cite{carlson07, afanasev07, blunden09}: a
small contribution at small scattering angles, corresponding to the virtual
photon polarization parameter $\varepsilon$=1, and a larger contribution for
small $\varepsilon$ values, and that the contributions become larger at large
$Q^2$ values.  This is consistent with the fact that the positron comparisons,
which were used to set upper limits on TPE effect, were typically focussed on
large $\varepsilon$ or low $Q^2$ values~\cite{arrington04b}.  The calculations
typically induce some non-linearity in the $\varepsilon$ dependence of the
reduced cross section~\cite{afanasev05a, blunden05a, abidin08}, which is
linear in the Born approximation, but the limits on nonlinearities in the
data~\cite{tvaskis06}, while significantly improved by the recent JLab
measurements~\cite{qattan05}, are not yet tight enough to be at odds with the
calculations.

It has been shown that the hadronic calculations of TPE~\cite{blunden05a}
can resolve the discrepancy up to 2--3~GeV$^2$, allowing for extraction of the
proton form factors~\cite{arrington07c} that includes an estimate of the
uncertainties for additional TPE effects at high $Q^2$.  However, this assumes
that TPE corrections fully explain the discrepancy.  If the discrepancy is
related to something else, such as higher order contributions to the radiative
corrections~\cite{afanasev07, weissbach09}, then the constraints applied,
based on the assumption that only TPE corrections are missing, could be
incorrect.  It is therefore critical to verify that TPE corrections fully
explain the discrepancy between Rosenbluth and polarization extractions of the
proton form factors.

In addition, it is important to remember that TPE contributions contribute to
all electromagnetic scattering processes.  It is generally assumed that these
corrections are small in almost all cases, and typically within the assumed
uncertainties applied for radiative corrections.  At the moment, we have no
way to verify this other than to make theoretical estimates of the TPE
contributions to other processes, and thus it is important to constrain these
calculations as well as possible in the case of elastic electron--proton
scattering, where there are multiple measurements that can be used to
quantitatively test the calculations.  While the focus has been on high
$Q^2$, it is also important to keep in mind that the TPE corrections do not
appear to be negligible at low $Q^2$, and thus the next generation of extremely
high precision measurements made at low $Q^2$ will also need better knowledge
of TPE corrections.

\section{Future positron-electron comparisons}

It is clear from the recent activity that obtaining a more complete
understanding of two-photon exchange effects is a matter of great interest and
importance (see Refs.~\cite{carlson07, afanasev07, blunden09} for details on
the theoretical and experimental activities). There are still quantitative
differences between different calculations, and it is crucial to determine the
reliability of the different approaches in their kinematic regions of
applicability, both to have complete confidence in our knowledge of the form
factors, but also to have reliable approaches that can be used to evaluate TPE
corrections for other reactions.

While recent experiments are attempting to examine TPE through
more detailed comparisons of the angular dependence of polarization and
cross section measurements, these can only constrain the effects of TPE, they
cannot isolate TPE contributions.  Other measurements, specifically of
polarization observables that are identically zero in the Born approximation,
can isolate TPE contributions.  However, these observables relate to the
imaginary part of the TPE amplitude, while the extractions of the form
factors are modified by the real part. Thus, these are important in evaluating
calculations of TPE corrections, but do not directly measure the effect on the
form factor extractions.

Comparisons of e$^+$--proton and e$^-$--proton scattering (as well
as $\mu^+$--p and $\mu^-$--p) have been used to set limits on TPE effects.
These contributions come through the interference of the one-photon and the
two-photon exchange amplitudes, and while the Born cross section is independent
of lepton charge, the interference term changes changes sign for positrons.
Thus, the comparison of e$^+$--p and e$^-$--p scattering isolates the TPE 
contribution (after correcting for the interference between electron and
proton bremsstrahlung, which also changes sign).  Previous
comparisons were interpreted as limiting the TPE contributions to the e--p
cross section at or below the 1\% level except at large $Q^2$~\cite{mar68}. 
However, due the the low luminosity of the secondary positron beams, the only
measurements above 2~GeV$^2$ were at small scattering angles, corresponding
to large $\varepsilon$ values.  A reexamination of the positron measurements,
in light of the form factor discrepancy between cross section and polarization
measurements showed that there was evidence for a $\varepsilon$ dependent
TPE correction~\cite{arrington04b}.  While the data were qualitatively
consistent with the TPE corrections necessary to explain the discrepancy, the
observed effect was only three sigma from zero, and the data at low
$\varepsilon$, where the TPE contributions were visible, was at low
$Q^2$ ($<1$~GeV$^2$), where the TPE effects are expected to be smaller.

Further measurements are required to adequately understand the impact of
TPE effects.  To be confident in our extraction of the form factors, high
$Q^2$ data are necessary to verify that TPE effects can fully explain the
discrepancy.  Mapping out the TPE contributions in detail will allow for
precise corrections in the low $Q^2$ region, where many high-precision
measurements are performed, and will also allow for detailed evaluations of
the TPE calculations.  Finally, with a high quality positron beams, direct
measurements of TPE effects in reactions beyond elastic e--p scattering will
become possible.

In the short term, there are three experiments planned to examine TPE effects
using positron beams.  A measurement at Novosibirsk~\cite{vepp_proposal} will
make a single high precision comparison of electron and positron scattering
at $Q^2=1.6$~GeV$^2$, $\varepsilon=0.4$.  The OLYMPUS experiment~\cite{kohl09}
will relocate the BLAST detector from MIT-BATES to the DORIS electron/positron
storage ring at DESY.  The experiment will be able to map out the epsilon
dependence in more detail using 2 GeV lepton beams, reaching a maximum $Q^2$
of 2.3~GeV$^2$ at $\varepsilon=0.35$ (and lower $Q^2$ for higher
$\varepsilon$ values).  Both of these experiments have clean lepton
beams but are limited by the total luminosity, even with large solid angle
detectors.  Nonetheless, they provide a dramatic improvement over the previous
measurements that included large scattering angle.  The last experiment
uses a mixed beam of positrons and electrons with a wide energy range, and
the large acceptance CLAS detector in Hall B at JLab is then used to detect both the scattered
lepton and struck proton, thus allowing for reconstruction of the charge and
energy of the incoming lepton~\cite{weinstein09}.  This will allow for
extraction of the TPE contributions over a range in $Q^2$, covering
approximately 0.5--2.0~GeV$^2$.  In this case, the luminosity is limited
by background rates in the detectors, and the $Q^2$ coverage may be
increased if modified shielding configurations are sufficient to reduce
these rates.  This experiment (JLab E07-005) provides broad kinematic coverage
and is the only one planned that can map out the TPE contributions at low
$Q^2$.  However, it requires the large acceptance and moderate resolution of
the CLAS spectrometer to fully reconstruct the events and to allow for control
of the systematics, and due to the rate limitations, it is difficult to know
exactly how high in $Q^2$ the data will extend.

These planned experiments will go a long way in verifying that TPE
contributions are responsible for the form factor discrepancy.  They will also
provide the first quantitative measure of TPE effects in the elastic e--p
cross section at low $\varepsilon$ and $Q^2>1$~GeV$^2$, were the effects are
believed to be most important.  However, further measurements will be 
important in fully understanding TPE effects.  The TPE calculations at higher
$Q^2$ are significantly less well constrained, and information on both the
scale and the $\varepsilon$-dependence at larger $Q^2$ values will be very
important.  In addition, a well defined positron beam of high luminosity would
allow for a survey of TPE contributions on a range of exclusive reactions.
Depending on the luminosity available, such measurements may be limited to low
$Q^2$, but this is the region where the majority of high precision
measurements are performed, and constraints on TPE contributions will be
most important.

A high quality positron beam at Jefferson Lab would allow for significant
extensions to the program of TPE studies, as well as related effects such as
Coulomb distortion~\cite{solvignon09}.  The main limitations of the planned
measurements are the luminosity, combined with the fact that the experiments
need to detect both the scattered lepton and struck proton in order to fully
reconstruct the event and sufficiently eliminate backgrounds.  A positron
beam with a small energy spread ($10^{-3}$ or better), coupled with a high
resolution spectrometer, would allow for a clean separation of the elastic
events detecting only the lepton or proton.  Proton-only detection, as used by
the ``Super-Rosenbluth'' experiment in Hall A~\cite{qattan05} has several
advantages in this case. Since only the proton is detected, the spectrometer
does not need to change polarity when the beam charge changes.  In addition,
low $\varepsilon$ values correspond to small scattering angles for the proton,
making it easier to access small $\varepsilon$ values and
providing a factor of 10--20 increase in the effective solid angle at low
$\varepsilon$ compared to lepton detection.  High $Q^2$ Rosenbluth separations
at SLAC~\cite{andivahis94} used beam currents up to 10~$\mu$A on a 15~cm liquid
hydrogen target to extract the form factors up to 7~GeV$^2$.  A measurement
using the HRS or HMS/SHMS spectrometers in Hall A or C would gain a factor of
5--10 in solid angle and 10--20 in cross section when detecting
protons at low $\varepsilon$, and thus could perform similar measurements
using positrons with a 100~nA positron beam, and even with 10~nA could make
measurements up to 5~GeV$^2$.

For a direct comparison of e$^+$ and e$^-$ scattering, it would be
beneficial to be able to change the beam fairly quickly.
However, in this case, one can make precision Rosenbluth separations
independently for positrons and electrons, using the proton detection technique
which minimizes the uncertainties on the $\varepsilon$ dependence of the
reduced cross section.  Therefore, one can make a direct comparison of the
$\varepsilon$ dependence extracted from electron and positron scattering,
rather than a direct comparison of individual cross sections.  In addition, the
TPE contributions go to zero for $\varepsilon \to 1$ ($\theta_e \to 0$), and
this can be used for a relative normalization if the data extend close enough
to $\varepsilon=1$ and the TPE corrections are sufficiently well behaved in
this region.

A beam of 10--100~nA would also allow for significant measurements using CLAS
in Hall B.  After the 12 GeV upgrade, the acceptance (and electron
identification) are limited at large scattering angles.  Thus, it is not as
well suited to looking for the large angle TPE expected in elastic e--p
scattering.  However, as 10--100~nA are typical operating currents in Hall B,
and the acceptance is very large, one could make a comparison of electron and
positron scattering simultaneously for a large number of exclusive reactions,
or use more specific trigger configurations to pick out specific reactions
with a lower cross section if models suggest that particular reactions
will be more sensitive to TPE contributions.  Again, it will be necessary to
carefully normalize the electron data to the positron data, taking multiple
configurations, \textit{e.g.}, positron data with same polarity as electron
data and with opposite polarity, to help minimize the systematics in the
comparisons of the results.  A quick change between positrons and electrons
would again be useful, but use of elastic scattering, after TPE corrections
are mapped out in detail, can be used as a check on differences in efficiency
between periods of positron and electron running.

In all of this, it will be important to include low $Q^2$ measurements. 
Calculations looking at the low $Q^2$ region~\cite{blunden05a, borisyuk06b}
suggest that the TPE correction goes to zero somewhere in the vicinity of
$Q^2=0.3$~GeV$^2$, and then changes sign and grows with decreasing $Q^2$. As
the low $Q^2$ region is where high precision extractions of the cross section
impact other observables, \textit{e.g.} the extraction of the strangeness
contribution to the nucleon form factors~\cite{tjon09, arrington07b}, and the
low $Q^2$ form factors that go into corrections of atomic hyperfine
splitting~\cite{carlson08}, precise limits are especially important in this
region.  If higher currents are available, one could also consider making
measurements of polarization observables in elastic e--p scattering.  The best
extractions of the form factors come from combining Rosenbluth and
polarization data, and at low $Q^2$, the TPE corrections on polarization
observables are small but not necessarily negligible.  With a polarized
positron beam, it would be able to make such measurements using a polarized
targets even for relatively low currents.  Such measurements would likely be
limited to larger $\varepsilon$ values, where one expects the TPE
contributions to be smaller.  This would suffice for extracting the
corrections to polarization observables, as a high-precision measure of the
asymmetries can be performed. To use this as a more detailed test of the TPE
calculations, high polarization and higher beam currents, probably at least
100~nA, would be required.

In summary, a great deal could be done to improve our understanding of the
two-photon exchange contributions, and thus the precision with which
we can extract the proton form factors with a positron beam at
Jefferson Lab.  An unpolarized beam of 10~nA would allow for significant 
progress over the existing and planned measurements of electron-positron
comparisons, and provide a first direct way to study TPE contributions in other
reactions.  If currents of 50--100~nA are available, these studies
could be dramatically expanded: high $Q^2$ measurements on the proton,
first measurements on the neutron, and better kinematic coverage for other
exclusive reactions on the proton.  These would dramatically improve our
tests of the TPE calculations that are necessary if we want significantly
improved precision on the next generation of electron-scattering experiments.
Finally, polarized beams would allow much independent tests of the details
of the TPE calculations, as well as providing direct measurements or
significant constraints on the impact of TPE on polarization measurements.

%%%%%%%%%%%%%%%%%%%%%%%%%%%%%%%%%%%%%%%%%%%%%%%%
%% BACKMATTER
%%%%%%%%%%%%%%%%%%%%%%%%%%%%%%%%%%%%%%%%%%%%%%%%

%\begin{theacknowledgments}
This work was supported by the U.S. Department of Energy, Office of Nuclear
Physics, under contract DE-AC02-06CH11357.
%\end{theacknowledgments}

%%%%%%%%%%%%%%%%%%%%%%%%%%%%%%%%%%%%%%%%%%%%%%%%
%% The bibliography can be prepared using the BibTeX program or
%% manually.
%%
%% The code below assumes that BibTeX is used.  If the bibliography is
%% produced without BibTeX comment out the following lines and see the
%% aipguide.pdf for further information.
%%
%% For your convenience a manually coded example is appended
%% after the \end{document}
%%%%%%%%%%%%%%%%%%%%%%%%%%%%%%%%%%%%%%%%%%%%%%%%

%%%%%%%%%%%%%%%%%%%%%%%%%%%%%%%%%%%%%%%%%%%%%%%%
%% You may have to change the BibTeX style below, depending on your
%% setup or preferences.
%%
%%
%% For The AIP proceedings layouts use either
%%%%%%%%%%%%%%%%%%%%%%%%%%%%%%%%%%%%%%%%%%%%

\bibliographystyle{aipproc_mod}   % if natbib is available
%\bibliographystyle{aipprocl} % if natbib is missing

%%%%%%%%%%%%%%%%%%%%%%%%%%%%%%%%%%%%%%%%%%%
%% You probably want to use your own bibtex database here
%%%%%%%%%%%%%%%%%%%%%%%%%%%%%%%%%%%%%%%%%%%
\bibliography{positron_tpe}

\begin{thebibliography}{48}
\expandafter\ifx\csname natexlab\endcsname\relax\def\natexlab#1{#1}\fi
\providecommand{\enquote}[1]{``#1''}
\expandafter\ifx\csname url\endcsname\relax
  \def\url#1{\texttt{#1}}\fi
\expandafter\ifx\csname urlprefix\endcsname\relax\def\urlprefix{URL }\fi
\providecommand{\eprint}[2][]{\url{#2}}

\bibitem[Arrington et~al.(2007{\natexlab{a}})]{arrington07a}
J.~Arrington, C.~D. Roberts, and J.~M. Zanotti, \emph{J. Phys.} \textbf{G34},
  S23--S52 (2007{\natexlab{a}}).

\bibitem[Perdrisat et~al.(2007)]{perdrisat07}
C.~F. Perdrisat, V.~Punjabi, and M.~Vanderhaeghen, \emph{Prog. Part. Nucl.
  Phys.} \textbf{59}, 694--764 (2007).

\bibitem[Miller(2003)]{miller03}
G.~A. Miller, \emph{Phys. Rev. C} \textbf{68}, 022201 (2003).

\bibitem[Kelly(2002)]{kelly02}
J.~J. Kelly, \emph{Phys. Rev. C} \textbf{66}, 065203 (2002).

\bibitem[Miller(2007)]{miller07}
G.~A. Miller, \emph{Phys. Rev. Lett.} \textbf{99}, 112001 (2007).

\bibitem[Miller and Arrington(2008)]{miller08b}
G.~A. Miller, and J.~Arrington, \emph{Phys. Rev. C} \textbf{78}, 032201 (2008).

\bibitem[Miller and Arrington(2009)]{miller09a}
G.~A. Miller, and J.~Arrington  (2009), \eprint{arXiv:0903.1617}.

\bibitem[Rosenbluth(1950)]{rosenbluth50}
M.~N. Rosenbluth, \emph{Phys. Rev.} \textbf{79}, 615 (1950).

\bibitem[Andivahis et~al.(1994)]{andivahis94}
L.~Andivahis, et~al., \emph{Phys. Rev. D} \textbf{50}, 5491 (1994).

\bibitem[Walker et~al.(1994)]{walker94}
R.~C. Walker, et~al., \emph{Phys. Rev. D} \textbf{49}, 5671 (1994).

\bibitem[Christy et~al.(2004)]{christy04}
M.~E. Christy, et~al., \emph{Phys. Rev. C} \textbf{70}, 015206 (2004).

\bibitem[Akhiezer and Rekalo(1968)]{akhiezer68}
A.~I. Akhiezer, and M.~P. Rekalo, \emph{Sov. Phys. Dokl.} \textbf{13}, 572
  (1968).

\bibitem[Dombey(1969)]{dombey69}
N.~Dombey, \emph{Rev. Mod. Phys.} \textbf{41}, 236 (1969).

\bibitem[Jones et~al.(2000)]{jones00}
M.~K. Jones, et~al., \emph{Phys. Rev. Lett.} \textbf{84}, 1398 (2000).

\bibitem[Gayou et~al.(2002)]{gayou02}
O.~Gayou, et~al., \emph{Phys. Rev. Lett.} \textbf{88}, 092301 (2002).

\bibitem[Arrington(2003)]{arrington03a}
J.~Arrington, \emph{Phys. Rev. C} \textbf{68}, 034325 (2003).

\bibitem[Qattan et~al.(2005)]{qattan05}
I.~A. Qattan, et~al., \emph{Phys. Rev. Lett.} \textbf{94}, 142301 (2005).

\bibitem[Mo and Tsai(1969)]{mo69}
L.~W. Mo, and Y.-S. Tsai, \emph{Rev. Mod. Phys.} \textbf{41}, 205--235 (1969).

\bibitem[Greenhut(1969)]{greenhut69}
G.~K. Greenhut, \emph{Phys. Rev.} \textbf{184}, 1860 (1969).

\bibitem[Mar et~al.(1968)]{mar68}
J.~Mar, et~al., \emph{Phys. Rev. Lett.} \textbf{21}, 482--484 (1968).

\bibitem[Maximon and Tjon(2000)]{maximon00}
L.~C. Maximon, and J.~A. Tjon, \emph{Phys. Rev. C} \textbf{62}, 054320 (2000).

\bibitem[Guichon and Vanderhaeghen(2003)]{guichon03}
P.~A.~M. Guichon, and M.~Vanderhaeghen, \emph{Phys. Rev. Lett.} \textbf{91},
  142303 (2003).

\bibitem[Blunden et~al.(2005)]{blunden05a}
P.~G. Blunden, W.~Melnitchouk, and J.~A. Tjon, \emph{Phys. Rev. C} \textbf{72},
  034612 (2005).

\bibitem[Afanasev et~al.(2005)]{afanasev05a}
A.~V. Afanasev, S.~J. Brodsky, C.~E. Carlson, Y.-C. Chen, and M.~Vanderhaeghen,
  \emph{Phys. Rev. D} \textbf{72}, 013008 (2005).

\bibitem[Borisyuk and Kobushkin(2007)]{borisyuk06b}
D.~Borisyuk, and A.~Kobushkin, \emph{Phys. Rev. C} \textbf{75}, 038202 (2007).

\bibitem[Borisyuk and Kobushkin(2008)]{borisyuk08}
D.~Borisyuk, and A.~Kobushkin, \emph{Phys. Rev. C} \textbf{78}, 025208 (2008).

\bibitem[Jain et~al.(2008)]{jain08}
P.~Jain, S.~D. Joglekar, and S.~Mitra, \emph{Eur. Phys. J.} \textbf{C57},
  671--680 (2008).

\bibitem[Kuhn and Weigel(2008)]{kuhn08}
M.~Kuhn, and H.~Weigel, \emph{Eur. Phys. J.} \textbf{A38}, 295--306 (2008).

\bibitem[Arrington and Sick(2004)]{arrington04c}
J.~Arrington, and I.~Sick, \emph{Phys. Rev. C} \textbf{70}, 028203 (2004).

\bibitem[Dong et~al.(2006)]{dong06}
Y.~B. Dong, C.~W. Kao, S.~N. Yang, and Y.~C. Chen, \emph{Phys. Rev. C}
  \textbf{74}, 064006 (2006).

\bibitem[Kondratyuk and Blunden(2006)]{kondratyuk06}
S.~Kondratyuk, and P.~G. Blunden, \emph{Nucl. Phys.} \textbf{A778}, 44--52
  (2006).

\bibitem[Afanasev and Carlson(2005)]{afanasev05b}
A.~V. Afanasev, and C.~E. Carlson, \emph{Phys. Rev. Lett.} \textbf{94}, 212301
  (2005).

\bibitem[Rekalo and Tomasi-Gustafsson(2004)]{rekalo04}
M.~P. Rekalo, and E.~Tomasi-Gustafsson, \emph{Eur. Phys. J.} \textbf{A22}, 331
  (2004).

\bibitem[Arrington and Sick(2007)]{arrington07b}
J.~Arrington, and I.~Sick, \emph{Phys. Rev. C} \textbf{76}, 035201 (2007).

\bibitem[Tjon et~al.(2009)]{tjon09}
J.~A. Tjon, P.~G. Blunden, and W.~Melnitchouk  (2009),
  \eprint{arXiv:0903.2759}.

\bibitem[Carlson and Vanderhaeghen(2007)]{carlson07}
C.~E. Carlson, and M.~Vanderhaeghen, \emph{Ann. Rev. Nucl. Part. Sci.}
  \textbf{57}, 171--204 (2007).

\bibitem[Afanasev(2007)]{afanasev07}
A.~V. Afanasev  (2007), \eprint{arXiv:0711.3065}.

\bibitem[Blunden(2009)]{blunden09}
P.~Blunden (2009), {\textit{Two-photon exchange: theoretical issues}, this
  volume}.

\bibitem[Arrington(2004)]{arrington04b}
J.~Arrington, \emph{Phys. Rev. C} \textbf{69}, 032201(R) (2004).

\bibitem[Abidin and Carlson(2008)]{abidin08}
Z.~Abidin, and C.~E. Carlson, \emph{Phys. Rev. D} \textbf{77}, 037301 (2008).

\bibitem[Tvaskis et~al.(2006)]{tvaskis06}
V.~Tvaskis, et~al., \emph{Phys. Rev. C} \textbf{73}, 025206 (2006).

\bibitem[Arrington et~al.(2007{\natexlab{b}})]{arrington07c}
J.~Arrington, W.~Melnitchouk, and J.~A. Tjon, \emph{Phys. Rev. C} \textbf{76},
  035205 (2007{\natexlab{b}}).

\bibitem[Weissbach et~al.(2009)]{weissbach09}
F.~Weissbach, K.~Hencken, D.~Trautmann, and I.~Sick  (2009),
  \eprint{arXiv:0903.0309}.

\bibitem[Arrington et~al.(2004)]{vepp_proposal}
J.~Arrington, D.~M. Nikolenko, et~al., Proposal for positron measurement at
  VEPP-3 (2004), \eprint{nucl-ex/0408020}.

\bibitem[Kohl(2009)]{kohl09}
M.~Kohl (2009), {\textit{OLYMPUS @ DESY: A proposal to definitively determine
  the contribution of multiple photon exchange in elastic lepton-nucleon
  scattering}, this volume}.

\bibitem[Weinstein(2009)]{weinstein09}
L.~Weinstein (2009), {\textit{Electron- and positron-proton elastic scattering
  in CLAS}, this volume}.

\bibitem[Solvignon(2009)]{solvignon09}
P.~Solvignon (2009), {\textit{Coulomb distortion in the inelastic regime}, this
  volume}.

\bibitem[Carlson et~al.(2008)]{carlson08}
C.~E. Carlson, V.~Nazaryan, and K.~Griffioen  (2008), \eprint{arXiv:0805.2603}.

\end{thebibliography}

%%%%%%%%%%%%%%%%%%%%%%%%%%%%%%%%%%%%%%%%%%%
%% Just a reminder that you may have to run bibtex
%% All of it up to \end{document} can be removed
%% if you don't like the warning.
%%%%%%%%%%%%%%%%%%%%%%%%%%%%%%%%%%%%%%%%%%%
\IfFileExists{\jobname.bbl}{}
 {\typeout{}
  \typeout{******************************************}
  \typeout{** Please run "bibtex \jobname" to optain}
  \typeout{** the bibliography and then re-run LaTeX}
  \typeout{** twice to fix the references!}
  \typeout{******************************************}
  \typeout{}
 }

\end{document}